\documentclass[oribibl]{llncs}
\usepackage[utf8]{inputenc}
\usepackage[english]{babel} 
\usepackage{multirow}
\usepackage{amssymb, amsmath, amsfonts,paralist}

\usepackage[ligature,inference]{semantic}
\usepackage{listings}
\usepackage{graphicx}

\usepackage{setspace}
\usepackage{multicol}
\usepackage{multirow}
\begin{document}

\title{Enforcing Secure Object Initialization in Java}

\author{Laurent Hubert\inst{1} \and Thomas Jensen\inst{2} \and Vincent
  Monfort\inst{2} \and David Pichardie\inst{2}}

\institute{CNRS/IRISA, France \and INRIA Rennes - Bretagne Atlantique/IRISA,
  France}
\maketitle

\begin{abstract}
  Sun and the CERT recommend for secure Java development to \emph{not allow
    partially initialized objects to be accessed}.  The CERT considers the
  severity of the risks taken by not following this recommendation as
  \emph{high}.  The solution currently used to enforce object initialization is
  to implement a coding pattern proposed by Sun, which is not formally checked.
  We propose a modular type system to formally specify the initialization policy
  of libraries or programs and a type checker to statically check at load time
  that all loaded classes respect the policy.  This allows to prove the absence
  of bugs which have allowed some famous privilege escalations in Java.  Our
  experimental results show that our safe default policy allows to prove 91\% of
  classes of \texttt{java.lang}, \texttt{java.security} and
  \texttt{javax.security} safe without any annotation and by adding 57 simple
  annotations we proved all classes but four safe.  The type system and its
  soundness theorem have been formalized and machine checked using Coq.
\end{abstract}




\newcommand{\remarques}[1]{}

\newcommand{\raw}{\textnormal{raw}}
\newcommand{\fram}[1]{\langle #1 \rangle}
\newcommand{\init}{\texttt{<init>}}
\newcommand{\jmp}{jmp}
\newcommand{\lookup}{\textnormal{lookup}}
\newcommand{\SetInit}{\texttt{SetInit}}
\newcommand{\AssertStaticInit}{\texttt{Raw}}
\newcommand{\AssertVirtualInit}{\texttt{Init}}
\sloppy{}

\newcommand{\beforefigurecaption}{\vspace*{-0.25em}}
\newcommand{\afterfigurecaption}{\vspace*{-.25em}}
\newcommand{\beforetablecaption}{\vspace*{0em}}
\newcommand{\aftertablecaption}{\vspace*{-1em}}

\section{Introduction}
\label{sec:rt-introduction}





The initialization of an information system is usually a critical
phase where essential defense mechanisms are being installed and a
coherent state is being set up. In object-oriented software, granting
access to partially initialized objects is consequently a delicate
operation that should be avoided or at least closely monitored.
Indeed, the CERT recommendation for secure Java
development~\cite{cert_sun_secure_coding_standard} clearly requires to
\emph{not allow partially initialized objects to be accessed}
(guideline OBJ04-J).  The CERT has assessed the risk if this
recommendation is not followed and has considered the severity as
\emph{high} and the likelihood as \emph{probable}%
.  They consider this recommendation as a first priority on a scale of three
levels.

The Java language and the Java Byte Code Verifier (BCV) enforce some properties
on object initialization, \emph{e.g.} about the order in which constructors of
an object may be executed, but they do not directly enforce the CERT
recommendation.  Instead, Sun provides a guideline that enforces the
recommendation.  Conversely, failing to apply this guidelines may silently lead
to security breaches.  In fact, a famous attack~\cite{dean96:java_security} used
a partially initialized class loader for privilege elevation.

We propose a twofold solution:
\begin{inparaenum}[(i)]
\item a modular type system which allows to express the initialization
  policy of a library or program, \emph{i.e.} which methods may access
  partially initialized objects and which may not; and
\item a type checker, which can be integrated into the BCV, to statically check
  the program at load time.
\end{inparaenum}
To validate our approach, we have
\emph{formalized} our type system, \emph{machine checked} its soundness proof
using the Coq proof assistant, and
\emph{experimentally validated} our solution on a large number of classes from
Sun's Java Runtime Environment (JRE).

Section~\ref{sec:overview} overviews object initialization in Java and its impacts
on security.  Section~\ref{sec:our-solution} then informally presents our type
system, which is then formally described in Section~\ref{sec:rt-formalization}.
Section~\ref{sec:experimentations} finally presents the experimental results we
obtained on Sun's JRE.

\section{Related Work.}
\label{sec:rt-related-work}
\remarques{(done) The paper must follow the template provided by the conference
  (i.e. numbering of sections, sub-sections, font size).  Related work that is
  presented in the introduction section could be a section of its own. }

Object initialization has been studied from different points of view.
Freund and Mitchell~\cite{freund03:type_system_java_bytecode_journal} have
proposed a type system that formalizes and enforces the initialization
properties ensured by the BCV, which are not sufficient to ensure that no
partially initialized object is accessed.
\remarques{(done) p2, the discussion of related work is unclear, especially for
  [7,8].}  %
Unlike local variables, instance fields have a default value (\lstinline!null!,
\lstinline!false! or \lstinline!0!) which may be then replaced by the program.
The challenge is then to check that the default value has been replaced before
the first access to the field (\emph{e.g.} to ensure that all field reads return
a non-null value).  This is has been studied in its general form by F\"ahndrich
and Xia~\cite{fahndrich07:delayed_types}, and Qi and
Myers~\cite{qi09:masked_types}.  
Those works are focused on enforcing invariants on fields and finely tracks the
different fields of an object.  They also try to follow the objects after their
construction to have more information on initialized fields.  This is an
overkill in our context.
Unkel and Lam studied another property of object initialization: stationary
fields~\cite{unkel08:infererence_stationary_fields}.  A field may be stationary
if all its reads return the same value.  There analysis also track fields of
objects and not the different initialization of an object.  In contrast to our
analysis, they stop to track any object stored into the heap. 
\\
Other work have targeted the order in which methods are called.  It has been
studied in the context of rare events (\emph{e.g.} to detect anomaly, including
intrusions).  We refer the interested reader to the survey of Chandola \emph{et
  al.}~\cite{chandola09:anomaly_detection}.  They are mainly interested in the
order in which methods are called but not about the initialization status of
arguments.  While we guarantee that a method taking a fully initialized receiver
is called after its constructor, this policy cannot be locally expressed with an
order on method calls as the methods (constructors) which needs to be called on
a object to initialize it depends on the dynamic type of the object.

\section{Context Overview}
\label{sec:overview}

%
%
%
%

Fig.~\ref{fig:classloader-original} is an extract of class
\lstinline!ClassLoader! of SUN's JRE as it was before 1997.  The security policy
which needs to be ensured is that \lstinline!resolveClass!, a security sensitive
method, may be called only if the security check l.~5 has succeeded.
\begin{figure}[t]
  \centering
  {\footnotesize
    \begin{lstlisting}[numbers=left, numberstyle=\tiny, numbersep=5pt]
public abstract class ClassLoader {
  private ClassLoader parent;
  protected ClassLoader() {
    SecurityManager sm = System.getSecurityManager();
    if (sm != null) {sm.checkCreateClassLoader();}
    this.parent = ClassLoader.getSystemClassLoader();
  }
  protected final native void resolveClass(Class c);
}
    \end{lstlisting}}
  \beforefigurecaption{}
  \caption{Extract of the ClassLoader of Sun's JRE}
  \label{fig:classloader-original}
  \afterfigurecaption{}
\end{figure}
To ensure this security property, this code relies on the properties enforced
on object initialization by the BCV.

\subsubsection{Standard Java Object Construction.}
\label{sec:standard-object-construction}

In Java, objects are initialized by calling a class-specific constructor
which is supposed to establish an invariant on the newly created object.
The BCV enforces two properties related to
these constructors. These two properties are necessary but, as
we shall see, not completely sufficient to avoid security problems due
to object initialization.  
\begin{property}\label{sec:prop-bcv-all-constructors}\sl
  Before accessing an object,
  \begin{inparaenum}[(i)]
  \item a constructor of its dynamic type has been called and
  \item each constructor either calls another constructor of the same class or a
    constructor of the super-class on the object under construction, except for
    \lstinline!java.lang.Object! which has no super-class.
  \end{inparaenum}
\end{property}
This implies that 
at least one constructor of $C$ and of each super-class of $C$ is called: it is
not possible to bypass a level of constructor.  To deal with
exceptional behaviour during object construction, the BCV enforces another
property ---~concisely described in \emph{The Java Language
  Specification}~\cite{gosling05:JLS_3rd_edition}, Section 12.5, or implied by the
type system described in the JSR202~\cite{buckley06:jsr202}).
\begin{property}\sl
  If one constructor finishes abruptly, then the whole construction of the
  object finishes abruptly.
\end{property}
Thus, if the construction of an object finishes normally, then all constructors
called on this object have finished normally.  Failure to implement this
verification properly led to a famous attack~\cite{dean96:java_security} in
which it was exploited that if code such as %
\lstinline!try {super();} catch(Throwable e){}!  in a constructor is not
rejected by the BCV, then malicious classes can create security-critical classes
such as class loaders.

\subsubsection{Attack on the class loader and the patch from Sun.}
\label{sec:attack-class-loader-1}


However, even with these two properties enforced, it is not guaranteed that
uninitialized objects cannot be used.
In Fig.~\ref{fig:classloader-original}, if the check fails, the method
\lstinline!checkCreateClassLoader! throws an exception and therefore terminates
the construction of the object, but the garbage collector then call a
\lstinline!finalize()! method, which is an instance method and has the object to
be collected as receiver (cf. Section 12.6 of~\cite{gosling05:JLS_3rd_edition}).

An attacker could code another class that extends \lstinline!ClassLoader! and
has a \lstinline!finalize()! method.  If run in a right-restricted context,
\emph{e.g.}  an applet, the constructor of \lstinline!ClassLoader! fails and the
garbage collector then call the attacker's \lstinline!finalize! method.  The
attacker can therefore call the \lstinline!resolveClass! method on it, bypassing
the security check in the constructor and breaking the security of Java.





The initialization policy enforced the BCV is in fact too weak: when a method is
called on an object, there is no guarantee that the construction of an object
has been successfully run.  %
An ad-hoc solution to this problem is proposed by SUN~\cite{sun_guidelines} in
its Guideline 4-3 \emph{Defend against partially initialized instances of
  non-final classes}: adding a special Boolean field to each class for which the
developer wants to ensure it has been sufficiently initialized.  This field, set
to \lstinline!false! by default, should be private and should be set to
\lstinline!true! at the end of the constructor.  Then, every method that relies
on the invariant established by the constructor must test whether this field is
set to \lstinline!true! and fail otherwise.
If \lstinline!initialized! is \lstinline!true!, the construction of the object up to
the initialization of \lstinline!initialized! has succeeded.  Checking if
\lstinline!initialized! is \lstinline!true! allows to ensure that sensitive code is
only executed on classes that have been initialized up to the constructor of the
current class.
Fig.~\ref{fig:classloader-patched} shows the same extract as in
Fig.~\ref{fig:classloader-original} but with the needed instrumentation (this is
the current implementation as of JRE 1.6.0\_16).
\begin{figure}[t]
  \centering
  {\footnotesize
    \begin{lstlisting}[numbers=left, numberstyle=\tiny, numbersep=5pt]
public abstract class ClassLoader {
  private volatile boolean initialized;
  private ClassLoader parent;
  protected ClassLoader() {
    SecurityManager sm = System.getSecurityManager();
    if (sm != null) {sm.checkCreateClassLoader();}
    this.parent = ClassLoader.getSystemClassLoader();
    this.initialized = true;}
  private void check() {
    if (!initialized) {
      throw new SecurityException(
        "ClassLoader object not initialized");}}
  protected final void resolveClass(Class c){
    this.check();
    this.resolveClass0(c);}
  private native void resolveClass0(Class c);
}
    \end{lstlisting}}
  \beforefigurecaption{}
  \caption{Extract of the ClassLoader of Sun's JRE}
  \label{fig:classloader-patched}
  \afterfigurecaption{}
\end{figure}

Although there are some exceptions and some methods are designed to access
partially initialized objects (for example to initialize the object), most
methods should not access partially initialized objects.  Following the
remediation solution proposed in the CERT's recommendation or Sun's guideline
4-3, a field should be added to almost every class and most methods should start
by checking this field.  This is resource consuming and error prone because it
relies on the programmer to keep track of what is the semantic invariant,
without providing the adequate automated software development tools.  It may
therefore lead not to functional bugs but to security breaches, which are
harder to detect.  In spite of being known since 1997, this pattern is not always
correctly applied to all places where it should be. This has lead to security
breaches, see \emph{e.g.}, the Secunia Advisory SA10056~\cite{SecuniaSA10056}.

\section{The right way: a type system}
\label{sec:our-solution}

We propose a twofold solution: first, a way to specify the security policy which
is simple and modular, yet more expressive than a single Boolean field; second,
a modular type checker, which could be integrated into the BCV, to check that
the whole program respects the policy.

\subsection{Specifying an Initialization Policy with Annotations.}
\label{sec:initialization-policy}


We rely on Java annotations and on one instruction to specify our initialization
policy.  We herein give 
the grammar of the annotations we use.
{  \footnotesize
  \begin{displaymath}
    \begin{array}{rcl}
      \texttt{V\_ANNOT} & ::= & \texttt{@Init | @Raw | @Raw(CLASS)}\\
      \texttt{R\_ANNOT} & ::= & \texttt{@Pre(V\_ANNOT) | @Post(V\_ANNOT)}
    \end{array}
  \end{displaymath}
}
We introduce two main annotations: \lstinline!@Init!, which specifies that a
reference can only point to a fully initialized object or the \lstinline!null!
constant, and \lstinline!@Raw!, which specifies that a reference may point to a
partially initialized object.  A third annotation, \lstinline!@Raw(CLASS)!,
allows to precise that the object may be partially initialized but that all
constructors up to and including the constructor of \lstinline!CLASS! must have
been fully executed. \emph{E.g.}, when one checks that \lstinline!initialized!
contains \lstinline!true! in \lstinline!ClassLoader.resolveClass!, one checks
that the receiver has the type \lstinline!@Raw(ClassLoader)!.  The annotations
produced by the \lstinline!V_ANNOT! rule are used for fields, method arguments
and return values.
In the Java language, instance methods implicitly take another argument: a
receiver ---~reachable through variable \lstinline!this!.
We introduce a \lstinline!@Pre! annotation to specify the type of the receiver
at the beginning of the method.  Some methods, usually called from constructors,
are meant to initialize their receiver.  We have therefore added the possibility
to express this by adding a \lstinline!@Post! annotation for the type of the
receiver at the end of the method.  These annotations take as argument an
initialization level produced by the rule \lstinline!V_ANNOT!.

Fig.~\ref{fig:raw_motivations} shows an example of \texttt{@Raw} annotations.
Class \lstinline!Ex1A! has an instance field \lstinline!f!, a constructor and a
getter \lstinline!getF!.
\begin{figure}[t]
  \centering
  \begin{multicols}{2}
    \footnotesize
    \begin{lstlisting}[numbers=left, numberstyle=\tiny, numbersep=5pt]
class Ex1A {
    private Object f;
    Ex1A(Object o){
        securityManagerCheck()
        this.f = o;}
    @Pre(@Raw(Ex1A))
    getF(){return this.f;}
}
\end{lstlisting}
\begin{lstlisting}[numbers=left, firstnumber=9, numberstyle=\tiny, numbersep=5pt]
class Ex1B extends Ex1A{
    Ex1B(){
        super();
        ... = this.getF();
    }
}
\end{lstlisting}
\end{multicols}
\beforefigurecaption{}
  \caption{Motivations for \lstinline!Raw(CLASS)! annotations}
  \label{fig:raw_motivations}
\afterfigurecaption{}
\end{figure}
This getter requires the object to be initialized at least up to
\lstinline!Ex1A! as it accesses a field initialized in its constructor.  The
constructor of \lstinline!Ex1B! uses this getter, but the object is not yet
completely initialized: it has the type \lstinline!Raw(Ex1A)! as it has finished
the constructor of \lstinline!Ex1A! but not yet the constructor
\lstinline!Ex1B!.  If the getter had been annotated with \lstinline!@Init!  it
would not have been possible to use it in the constructor of \lstinline!Ex1B!.

Another part of the security policy is the \SetInit{} instruction, which mimics
the instruction \lstinline!this.initialized = true! in Sun's guideline.  It is
implicitly put at the end of every constructor but it can be explicitly placed
before.  It declares that the current object has completed its initialization up
to the current class.  Note that the object is not yet considered fully
initialized as it might be called as a parent constructor in a subclass.
\remarques{(done)why? because the constructor might in fact be called (as a
  parent constructor) in a subclass?} %
The instruction can be used, as in Fig.\ref{fig:setinit-example}, in a
constructor after checking some properties and before calling some other method.
\begin{figure}[t]
  \centering
{\footnotesize
\begin{lstlisting}[numbers=left, numberstyle=\tiny, numbersep=5pt]
public C() {
  ...
  securityManagerCheck(); // perform dynamic security checks 
  SetInit;                // declare the object initialized up C
  Global.register(this);  // the object is used with a method 
}                         // that only accept Raw(C) parameters
\end{lstlisting}
}\beforefigurecaption{}
  \caption{An Example with \SetInit{}}
  \label{fig:setinit-example}
\afterfigurecaption{}
\end{figure}

Fig.~\ref{fig:classloader-annotated} shows class \lstinline!ClassLoader! with
its policy specification.  The policy ensured by the current implementation of
Sun is slightly weaker: it does not ensure that the receiver is fully
initialized when invoking \lstinline!resolveClass! but simply checks that the
constructor of \lstinline!ClassLoader! has been fully run.
\remarques{(can we be more precise without repeating ourselves for the third
  time?)  It's not so clear precisely in what way the policy of Sun's current
  implementation is weaker.}%
On this example, we can see that the constructor has the annotations
\lstinline!@Pre(@Raw)!, meaning that the receiver may be completely
uninitialized at the beginning, and \lstinline!@Post(@Raw(ClassLoader))!,
meaning that, on normal return of the method, at least one constructor for each
parent class of \lstinline!ClassLoader!  and a constructor of
\lstinline!ClassLoader! have been fully executed.
\remarques{(done)For Fig 5, I was wondering if there is an implicit SetInit at
  the end of every constructor. I don't think there is, right?}
\begin{figure}[th]
  \centering
  {\footnotesize
    \begin{lstlisting}[numbers=left, numberstyle=\tiny, numbersep=5pt]
public abstract class ClassLoader {
  @Init private ClassLoader parent;
  @Pre(@Raw) @Post(@Raw(ClassLoader))
  protected ClassLoader() {
    SecurityManager sm = System.getSecurityManager();
    if (sm != null) {sm.checkCreateClassLoader();}
    this.parent = ClassLoader.getSystemClassLoader();
  }
  @Pre(@Init) @Post(@Init)
  protected final native void resolveClass(@Init Class c);
}
    \end{lstlisting}}
\beforefigurecaption{}
  \caption{Extract of the ClassLoader of Sun's JRE}
  \label{fig:classloader-annotated}
\afterfigurecaption{}
\end{figure}

We define as default values the most precise type that may be use in each
context.  This gives a \emph{safe by default} policy and lowers the burden of
annotating a program.\remarques{p7, the paragraph below fig 5 needs fixing.}
\begin{itemize}
\item Fields, method parameters and return values are fully initialized objects
  (written \lstinline!@Init!).
\item Constructors take a receivers uninitialized at the beginning
  (\lstinline!@Pre(@Raw)!) and initialized up-to the current class at the end
  (written \lstinline!@Post(@Raw(C))! if in the class \lstinline!C!).
\item Other methods take a receiver fully initialized (\lstinline!@Pre(@Init)!).
\item Except for constructors, method receivers have the same type at the end as
  at beginning of the method (written \lstinline!@Post(A)! if the method has the
  annotation \lstinline!@Pre(A)!).
\end{itemize}
If we remove from Fig.~\ref{fig:classloader-annotated} the default annotations,
we obtain the original code in Fig.~\ref{fig:classloader-original}.  It shows
that despite choosing the strictest (and safest) initialization policy as
default, the annotation burden can be kept low.

\subsection{Checking the Initialization Policy.}
\label{sec:checking-policy}


We have chosen static type checking for at least two reasons. %
Static type checking allows for more performances (except for some rare cases),
as the complexity of static type checking is linear in the \emph{code size},
whereas the complexity of dynamic type checking is linear in the \emph{execution
  time}.
Static type checking also improves reliability of the code:
if a code passes the type checking, then the code is correct with respect to its
policy, whereas the dynamic type checking only ensures the correction of a
particular execution.

Reflection in Java allows to retrieve code from the network or to dynamically
generates code.  Thus, the whole code may not be available before actually
executing the program.  Instead, code is made available class by class, and
checked by the BCV at linking time, before the first execution of each method.
As the whole program is not available, the type checking must be modular: there
must be enough information in a method to decide if this method is correct and,
if an incorrect method is found, there must exist a safe procedure to end the
program (usually throwing an exception), \emph{i.e.} it must not be too late.


To a have a modular type checker while keeping our security policy simple,
method parameters, respectively return values, need to be contra-variant,
respectively co-variant, \emph{i.e.} the policy of the overriding methods needs
to be at least as general as the policy of the overridden method.  Note that
this is not surprising: the same applies in the Java language (although Java
imposes the invariance of method parameters instead of the more general
contra-variance), and when a method call is found in a method, it allows to rely
on the policy of the resolved method (as all the method which may actually be
called cannot be known before the whole program is loaded).

\section{Formal Study of the Type System}
\label{sec:rt-formalization}

The purpose of this work is to provide a type system that enforces at
load time an important security property. The semantic soundness of
such mechanism is hence crucial for the global security of the Java
platform. In this section, we formally define the type system
and prove its soundness with respect to an operational semantics.
All the results of this section have been machine-checked with the Coq
proof assistant\footnote{The development can be downloaded 
at \url{http://www.irisa.fr/celtique/ext/rawtypes/}}.


\newcommand{\Expr}{\mathit{Expr}}
\newcommand{\Alloc}{\mathit{Alloc}}
\newcommand{\Var}{\mathit{Var}}
\newcommand{\type}{\mathit{Type}}
\newcommand{\Field}{\mathit{Field}}
\newcommand{\Meth}{\mathit{Meth}}
\newcommand{\Class}{\mathit{Class}}
\newcommand{\Power}{\mathcal{P}}
\newcommand{\prog}{\mathit{Prog}}
\mathlig{|->}{\mapsto}
\mathlig{|-}{\vdash}
\mathlig{<-}{\leftarrow}
\mathlig{->}{\rightarrow}
\mathlig{=>}{\implies}
\newcommand{\dasig}{\mathrel{\mathop{::}}=}
\newcommand{\fclasses}{\mathsf{classes}}
\newcommand{\ffields}{\mathsf{fields}}
\newcommand{\flookup}{\mathsf{lookup}}
\newcommand{\fmain}{\mathsf{main}}
\newcommand{\fsuper}{\mathsf{super}}
\newcommand{\fmethods}{\mathsf{methods}}
\newcommand{\finit}{\mathsf{init}}
\newcommand{\pto}{\rightharpoonup}
\newcommand{\finstrs}{\mathsf{instrs}}
\newcommand{\fhandler}{\mathsf{handler}}
\newcommand{\Exception}{\mathit{Exc}}
\newcommand{\fpre}{\mathsf{pre}}
\newcommand{\fpost}{\mathsf{post}}
\newcommand{\fret}{\mathsf{rettype}}
\newcommand{\farg}{\mathsf{argtype}}
\newcommand{\Init}{\mathit{Init}}
\newcommand{\Raw}{\mathit{Raw}}
\newcommand{\instr}{\mathit{Instr}}
\newcommand{\pp}{\mathcal{L}}
\newcommand{\ins}{\mathit{ins}}
\newcommand{\vnull}{\mathit{null}}
\newcommand{\varg}{\mathit{arg}}
\newcommand{\vif}{\mathit{if}}
\newcommand{\vsuper}{\mathit{super}}
\newcommand{\vreturn}{\mathit{return}}
\newcommand{\vnew}{\mathit{new}}
\newcommand{\vinit}{\mathit{init}}
\newcommand{\vthis}{\mathit{this}}
\newcommand{\vsetinit}{\mathit{SetInit}}
\newcommand{\Rawtop}{\mathit{Raw}^{\bot}}
\newcommand{\locations}{\mathbb{L}}
\newcommand{\objects}{\mathbb{O}}
\newcommand{\values}{\mathbb{V}}
\newcommand{\heap}{\mathbb{H}}
\newcommand{\np}{\mathsf{np}}
\newcommand{\states}{\mathbb{S}}
\newcommand{\locvar}{\mathbb{M}}
\newcommand{\emptyostack}{\varepsilon}
\newcommand{\state}[1]{\langle #1 \rangle}
\newcommand{\ee}{\overline{e}}

\subsubsection{Syntax}
\label{sec:rt-syntax}
\begin{figure}[t]
  \centering
  \begin{small}
$
 \begin{array}{rrrrl}
\multicolumn{5}{c}{
x,y,r \in \Var \quad
f \in \Field \quad
e \in \Exception \quad
i \in \pp = \mathbb{N}} \\
p&\in&\prog  & \dasig & \{
  \begin{array}[t]{l}
 \fclasses\in\Power(\Class),~
 \fmain\in\Class,\\
 \ffields\in\Field\to\type,~
 \flookup\in\Class\to\Meth\pto\Meth\}
\end{array}\\
c&\in&\Class  & \dasig & \{
      \fsuper \in \Class_\bot,~
      \fmethods \in \Power(\Meth),~
      \finit \in \Meth
      \} \\
m&\in&\Meth & \dasig & \{
\begin{array}[t]{l}
    \finstrs\in\instr~\mathit{array},~
    \fhandler\in\pp\to\Exception\to\pp_\bot,\\
    \fpre\in\type,~\fpost\in\type,~\farg\in\type,~\fret\in\type
   \}
 \end{array}
 \\
\tau&\in&\type  & \dasig & \Init \mid \Raw(c) \mid \Rawtop \\
e&\in&\Expr  & \dasig & \vnull \mid x \mid e.f\\
\ins&\in&\instr  & \dasig & x <- e \mid x.f <- y \mid 
x <- \vnew\ c(y) \mid \vif\ (\star)\ \jmp \mid \\
&&&& \vsuper(y) \mid x <- r.m(y) \mid \vreturn~x \mid \vsetinit\\
\end{array}
$
\end{small}
\beforefigurecaption{}
  \caption{Language Syntax.}
  \label{fig:syntax}
  \afterfigurecaption{}
\end{figure}

Our language is a simple language in-between Java source and Java bytecode.  Our
goal was to have a language close enough to the bytecode in order to easily
obtain, from the specification, a naive implementation at the bytecode level
while keeping a language easy to reason with.  It is based on the decompiled
language from Demange \emph{et al.}~\cite{DEMANGE:2009:INRIA-00414099:2} that
provides a stack-less representation of Java bytecode programs.
Fig.~\ref{fig:syntax} shows the syntax of the language.  A program is a record
that handles a set of classes, a main class, a type annotation for each field
and a lookup operator. This operator is used do determine during a virtual call
the method $(p.\flookup~c~m)$ (if any) that is the first overriding version of a
method $m$ in the ancestor classes of the class $c$.  A class is composed of a
super class (if any), a set of method and a special constructor method
$\finit$. A method handles an array of instructions, a handler function such
that $(m.\fhandler~i~e)$ is the program point (if any) in the method $m$ where
the control flows after an exception $e$ has been thrown at point $i$. Each
method handles also four initialization types for the initial value of the
variable \lstinline!this! ($m.\fpre$), its final value ($m.\fpost$), the type of
its formal parameter\footnote{For the sake of simplicity, each method has a
  unique formal parameter $\varg$.} ($m.\farg$) and the type of its return value
($m.\fret$).  The only expressions are the $\vnull$ constant, local variables
and field reads.  For this analysis, arithmetic needs not to be taken into
account.  We only manipulate objects.
The
instructions are the assignment to a local variable or to a field,
object creation ($\vnew$)\footnote{ Here, the same instruction
  allocates the object and calls the constructor. At bytecode level
  this gives raise to two separated instructions in the program
  (allocation and later constructor invocation) but the intermediate
  representation generator~\cite{DEMANGE:2009:INRIA-00414099:2} on
  which we rely is able to recover such construct.},
(non-deterministic) conditional jump, super constructor call, virtual
method call, return, and a special instruction that we introduce for
explicit object initialization: $\vsetinit$.


\subsubsection{Semantic Domains}

\begin{figure}[t]
  \centering
  \begin{small}
$
  \begin{array}{c}
  \begin{array}{rclcll} 
    \overline{\Exception} &\owns& \ee & ::= & e \mid \bot & \text{(exception flag)} \\
  \locations &\owns &l&& & {\text{(location)}}\\
  \values  &\owns & v & ::= & l \mid \vnull  & {\text{(value)}}\\
 \locvar = \Var \rightarrow \values &\owns &\rho && &{\text{(local variables)}}\\
    \objects = \Class \times \Class_\bot \times (\Field \rightarrow \values) &\owns& o
       &::=& [c,c_\vinit,o]  &{\text{(object)}}\\
\heap =  \locations \to \objects_\bot &\owns&\sigma&&&{\text{(heap)}} \\
    CS &\owns &cs &::=& (m,i,l,\rho,r)::cs \mid \emptyostack &{\text{(call stack)}}\\
  \states = \Meth\times\pp\times\locvar\times\heap\times CS\times \overline{\Exception}  &\owns&st
             &::=& \state{m,i,\rho,\sigma,cs}_{\ee}
             &{\text{(state)}}
             \\
  \end{array}
  \end{array}
  $
\end{small}\beforefigurecaption{}
  \caption{Semantic Domains.}
  \label{fig:domains}\afterfigurecaption{}
\end{figure}

Fig.~\ref{fig:domains} shows the concrete domain used to model the
program states.  The state is composed of the current method $m$, the
current program point $i$ in $m$ (the index of the next instruction to
be executed in $m.\finstrs$), a function for local variables, a heap,
a call stack and an exception flag. The heap is a partial function which
associates to a location an object $[c,c_\vinit,o]$ with $c$ its type,
$c_\vinit$ its current initialization level and $o$ a map from field to value
(in the sequel $o$ is sometimes confused with the object itself). An initialization
$c_\vinit\in\Class$  means that each constructors of $c_\vinit$
and its super-classes have been called on the object and have returned without abrupt
termination. The exception flag is used to handle exceptions: a state 
$\state{\cdots}_e$ with $e\in\Exception$ is reached after an exception $e$
has been thrown. The execution then looks for a handler in the current method and if
necessary in the methods of the current call stack. When equal to $\bot$, the flag is
omitted (normal state). The call stack records the program points of the pending calls together
with their local environments and the variable that will be assigned with the result of the
call. 

\subsubsection{Initialization types}

We can distinguish three different kinds of initialization
types. 
Given a heap $\sigma$ we define a value type judgment $h\vdash v:\tau$ between
values and types with the following rules.
\begin{small}
  \begin{center}
$
\inference{}{\sigma \vdash \vnull : \tau}
\inference{}{\sigma \vdash l : \Rawtop}    
\inference{\sigma(l)=[c_{\mathit{dyn}},c_\vinit,o] \\
 \forall c', c_{\mathit{dyn}} \preceq c' \land c \preceq c' \Rightarrow c_\vinit \preceq c'}{\sigma \vdash l : \Raw(c)}    
\inference{\sigma(l)=[c,c,o]}{\sigma \vdash l : \Init}    
$
  \end{center}
\end{small}
The relation $\preceq$ here denotes the reflexive transitive closure of
the relation induced by the $\vsuper$ element of each class.
$\Rawtop$ denotes a reference to an object which may be
completely uninitialized (at the very beginning of each constructor).
$\Init$ denotes a reference to an object which has been
completely initialized%
.  Between those two ``extreme'' types, a value may be typed as
$\Raw(c)$ if at least one constructor of $c$ and of each parent of $c$
has been executed on all objects that may be reference from this
value.  We can derive from this definition the sub-typing relation
$\Init \sqsubseteq \Raw(c) \sqsubseteq \Raw(c') \sqsubseteq \Rawtop$
if $c\preceq c'$. It satisfies the important monotony property
\begin{center}
  $\forall \sigma\in\heap,\forall v\in\values, \forall \tau_1,\tau_2\in\type,~ \tau_1\sqsubseteq \tau_2\land 
      \sigma |- v : \tau_1 \Rightarrow  \sigma |- v : \tau_2$
 \end{center}
 Note that the sub-typing judgment is disconnected from the static
 type of object. In a first approach, we could expect to manipulate a
 pair $(c,\tau)$ with $c$ the static type of an object and $\tau$ its
 initialization type and consider equivalent both types $(c,\Raw(c))$
 and $(c,\Init)$. Such a choice would however impact deeply on the
 standard dynamic mechanism of a JVM: each dynamic cast from $A$ to
 $B$ (or a virtual call on a receiver) would requires to check that an
 object has not only an initialization level set up to $A$ but also
 set up to $B$.

\begin{figure}[t]
  \centering
  \begin{small}
$
\begin{array}[c]{c}
\inference{ m.\finstrs[i] = x <- \vnew~c(y) & x\not=\vthis & \mathit{Alloc}(\sigma,c,l,\sigma')
& \sigma' \vdash \rho(y) : c.\finit.\farg}
{\state{m,i,\rho,\sigma,cs}\Rightarrow
 \state{c.\finit,0,[\cdot\mapsto\vnull][\vthis\mapsto l][\varg\mapsto\rho(y)],\sigma',(m,i,\rho,x)::cs}}
\\[1.5em]
\inference{ m.\finstrs[i] = \vsetinit& m = c.\finit & \rho(\vthis) = l & \mathit{SetInit}(\sigma,c,l,\sigma')}
{\state{m,i,\rho,\sigma,cs}\Rightarrow
 \state{m,i{+1},\rho,\sigma',cs}}
\\[1.5em]
\inference{ m.\finstrs[i] = \vreturn~x & \rho(\vthis) = l &
((\forall c,~m\not=c.\finit) \Rightarrow \sigma=\sigma')\\
(\forall c,~m=c.\finit \Rightarrow \mathit{SetInit}(\sigma,c,l,\sigma') \land x=\vthis)}
{\state{m,i,\rho,\sigma,(m',i',\rho',r)::cs}\Rightarrow
 \state{m',i'{+1},\rho'[r\mapsto \rho(x)],\sigma',cs}}
\end{array}
$
\end{small}\beforefigurecaption{}
  \caption{Operational Semantics (excerpt).}
  \label{fig:opsem}\afterfigurecaption{}
\end{figure}

\begin{figure}[!b]
\begin{small}
\centering
\begin{minipage}[t]{1.0\linewidth}
\bf {Expression typing}
\end{minipage}

$    \begin{array}{c}
      \inference
      {}
      {L |- e.f : (p.\ffields~f) }\qquad
      \inference
      {}
      {L |- x : L(x)}\qquad
      \inference
      {}
      {L |- \vnull : \Init}
    \end{array}$\vspace*{1ex}
\begin{minipage}[t]{1.0\linewidth}
\bf {Instruction typing}
\end{minipage}

$    \begin{array}{c}      
      \inference%
      {L |- e : \tau & x\not=\vthis}%
      {m |- x <- e : L -> L[x |-> \tau]} ~
      \inference%
      {L(y) \sqsubseteq (p.\ffields~f)}%
      {m |- x.f <- y : L -> L}~
      \inference
      {}
      {\Gamma,m |- \vif\star\ \jmp : L -> L}\\[1em]
      \inference%
      {L(\vthis) \sqsubseteq m.\fpost & 
        L(x) \sqsubseteq m.\fret & 
        (\forall c,~ m=c.\finit \Rightarrow L(\vthis) \sqsubseteq \Raw(c.\fsuper))}%
      {m |- \vreturn~x : L -> L}\\[1em]
      \inference%
      {L(y) \sqsubseteq c.\finit.\farg }%
      {m |- x <- \vnew~c(y) : L -> L[x|->\Init]}\qquad
      \inference%
      {c' = c.\vsuper & L(y) \sqsubseteq c'.\finit.\farg }%
      {c.\finit |- \vsuper(y) : L -> L[\vthis|->\Raw(c')]}\\[1em]
      \inference%
      {L(r) \sqsubseteq m.\fpre & 
       L(y) \sqsubseteq m.\farg & }%
      {m |-  x <- r.m'(y) : L -> L[r|->m.\fpost][x|->m.\fret]} ~
      \inference%
      {L(\vthis) \sqsubseteq \Raw(c.\fsuper)}%
      {c.\finit |- \vsetinit : L -> L}\\[1em]
    \end{array}$
  \end{small}
  \label{fig:typerul}
\beforefigurecaption{}
\caption{Flow sensitive type system}
\end{figure}

\subsubsection{Operational Semantics}
We define the operational semantics of our language as a small-step transition relation
over program states. A fixed program $p$ is implicit in the rest of this section.
Fig.~\ref{fig:opsem} presents some selected rules for this relation. 
The rule for the $\vnew$ instruction includes both the allocation and the
call to the constructor.  We use the auxiliary predicate
$\mathit{Alloc}(\sigma,c,l,\sigma')$ which allocate a fresh location
$l$ in heap $\sigma$ with type $c$, initialization type equals to $\bot$
and all fields set equal to $\vnull$. The constraint
$\sigma' \vdash \rho(y) : c.\finit.\farg$ explicitly asks the caller  
of the constructor to give a correct argument with respect to the policy of 
the constructor. Each call rules of the semantics have similar constraints. The execution
is hence stuck when an attempt is made to call a method with badly typed parameters.
The $\vsetinit$ instruction updates the initialization level of
the object in $\vthis$. It relies on the predicate 
$\mathit{SetInit}(\sigma,c,l,\sigma')$ which specifies that $\sigma'$ is a copy of 
$\sigma$ where the object at location $l$ has now the initialization tag set to $c$
if the previous initialization was $c.\vsuper$. It
 forces the current object (\lstinline!this!) to be
considered as initialized up to the current class (\emph{i.e.} as if the
constructor of the current class had returned, but not necessarily the
constructors of the subsequent classes).  This may be used in the constructor,
once all fields that need to be initialized have been initialized and if some
method requiring a non-raw object needs to be called.  Note that this
instruction is really sensitive: using this instruction too early in a
constructor may break the security of the application. 
The $\vreturn$ instruction uses the same predicate when invoked in a constructor. For convenience
we requires each constructor to end with a $\vreturn~\vthis$ instruction.

\subsubsection{Typing judgment} Each instruction $\ins$ of a method
$m$ is attached a typing rule (given in Fig.~\ref{fig:typerul}) $m
|- \ins : L -> L'$ that constraint the type of variable before ($L$)
and after ($L'$) the execution of $\ins$.

\begin{definition}[Well-typed Method]
  A method $m$ is \emph{well-typed} if there exists flow sensitive variable
  types $L\in\pp\to\Var\to\type$ such that
\begin{itemize}
\item $m.\fpre \sqsubseteq L(0,\vthis)$ and $m.\farg \sqsubseteq L(0,\varg)$,
\item for all instruction $\ins$ at point $i$ in $m$ and every
  successor $j$ of $i$, there exists a map of variable types
  $L'\in\Var\to\type$ such that $L'\sqsubseteq L(j)$ and the typing
  judgment $m\vdash \ins: L(i) \to L'$  holds. If $i$ is in the handler $j$ of an exception
 $e$ (\emph{i.e} $(m.\fhandler~i~e=j)$) then $L(i)\sqsubseteq L(j)$.
\end{itemize}
\end{definition}

The typability of a method can be decided by turning the set of typing rules
into a standard dataflow problem.  The approach is
standard~\cite{freund03:type_system_java_bytecode_journal} and not formalized
here.

\begin{definition}[Well-typed Program]
  A program $p$ is \emph{well-typed} if all its methods are well-typed and the
  following constraints holds:
  \begin{enumerate}
  \item for every method $m$ that is overridden by a method $m'$ (\emph{i.e} 
    there exists c, such that $(p.\flookup~c~m=~m')$),\\
\hspace*{12ex}$
\begin{array}[c]{cccc}
m.\fpre \sqsubseteq m'.\fpre &\land&   m.\farg \sqsubseteq m'.\farg &\land \\
   m.\fpost \sqsupseteq m'.\fpost &\land&   m.\fret \sqsupseteq m'.\fret
 \end{array}
 $
\item in each method, every first point, jump target and handler point
  contain an instruction and every instruction (except $\vreturn$) has a next instruction,
 \item the default constructor $c.\finit$ of each class $c$ is unique.
 \end{enumerate}
\end{definition}
In this definition only point 1 is really specific to the current type system. The other points are necessary
to established the progress theorem of the next section.

\subsubsection{Type soundness}
We rely on an auxiliary notion of well-formed states that capture the semantics
constraints enforce by the type system.  A state $\state{m,i,\rho,\sigma,cs}$ is
\emph{well-formed} (wf) if there exists a type annotation
$L_p\in(\Meth\times\pp)\to(\Var\to\type)$ such that
  \begin{center}
$
\begin{array}[c]{lr}
\forall l\in\pp, \forall o\in\objects, \sigma(l) = o \Rightarrow \sigma |- o(f) : (p.\ffields~f) & \text{(wf. heap)}\\
\forall x\in\Var, \sigma |- \rho(x) : L_p[m,i](x)  & \text{(wf. local variables)}\\
\forall (m',i',\rho',r)\in cs,~ \forall x, \sigma |- \rho'(x) : L_p[m',i'](x)  & \text{(wf. call stack)}\\
\end{array}
$
  \end{center}

Given a well-typed program $p$ we then
establish two key theorems. First, any valid transition from a
well-formed state leads to another well-formed state
(\emph{preservation}) and then, from every well-formed state there
exists at least a transition (\emph{progress}). As a consequence we
can establish that starting from an initial state (which is always
well-formed) the execution is never stuck, except on final
configuration. This ensures that all initialization constraints given
in the operational semantics are satisfied without requiring any dynamic
verification. 

\subsubsection{Limitations}
The proposed language has some limitations compared to the Java (bytecode)
language.  Static fields and arithmetic have not been introduced but are handled
by our implementation and do not add particular difficulties. Arrays have not
been introduced in the language neither.  Our implementation conservatively
handles arrays by allowing only writes of $\Init$ references in arrays.
Although this approach seems correct it has not been proved and it is not
flexible enough (cf. Section~\ref{sec:experimentations}).  Multi-threading as
also been left out of the current formalization but we conjecture the soundness
result still holds with respect to the Java Memory Model because of the flow
insensitive abstraction made on the heap.  As for the BCV, native methods may
brake the type system.  It is their responsibility to respect the policy
expressed in the program.

\newcommand{\comment}[1]{}
\comment{
The type system proposed in the previous sections is modular and allows the
type-checker to be integrated into the BCV, to statically verify at load time
that the whole program is correct.
For its integration, the type checker must be modular in the sens that it must
not need to load other classes to check the current method.  To avoid loading
other classes to check the current method \lstinline!C.m!, the policy of the
methods called from \lstinline!C.m! must be included in class \lstinline!C!.  Then,
when the method is resolved for the first time, the checker must check that copy
of the initialization policy included in the caller matches the initialization
policy of the resolved method.
Fields imposes the same requirements.
%
To avoid having to add the policy to each method call in every class, and to
ease the evaluation of our type system, we first have implemented a standalone
prototype.  When an invoke instruction is found, the method is resolved and the
security policy is found as an attribute of the method (see Sect 4.6 and 4.7
of~\cite{lindholm99:jvm_spec} or Section 4.7 and 4.8 of~\cite{buckley06:jsr202}).

As shown by the attack sketch in Section~\ref{sec:rt-introduction}, constructors
are not the only methods which may be invoked by default with uninitialized
object.  It is also the case with \lstinline!finalize! methods and with methods
used to deserialize~\footnote{\emph{Serialization} is the feature which allows
  to transmit object from an instance of one process to another ---~either with
  the same program later in time (as a storage feature) or with another
  concurrent process (as a communication feature).} objects:
\lstinline!readObject!, \lstinline!readObjectNoData! and \lstinline!readResolve!.
}

\section{A Case Study: Sun's JRE}
\label{sec:experimentations}
\remarques{(done)In Section 4, the discussion of special/unhandled cases is a bit
  confusing. It's unclear how the remaining 4 (=381-377) classes can be dealt
  with, if at all. Are these the ``unhandled'' cases discussed on page 15 (but
  there are only 3 of those?)

  The various categories of tricky cases should be clearer:
  \begin{itemize}
  \item cases where you need the dynamic cast, but which can be handled then
  \item one case with arrays
  \item some cases with inner classes
  \end{itemize}
  Could for any of the last two categories (which you cannot handle at all,
  right?) code be rewritten so that they could be handled? Or is some further
  extension allowing more flexible handling of arrays needed?}

In order to show that our type system allows to verify legacy code with only a
few annotations, we implemented a standalone prototype, handling the full Java
bytecode, and we tested all classes of packages \lstinline!java.lang!,
\lstinline!java.security!  and \lstinline!javax.security! of the JRE1.6.0\_20.

348 classes out of 381 were proven safe \emph{w.r.t.} the default policy without
any modification.  By either specifying the actual policy when the default
policy was too strict, or by adding cast instructions (see below) when the type
system was not precise enough, we were able to verify 377 classes, that is to
say 99\% of classes.
We discuss below the 4 remaining classes that are not yet proven correct by our
analysis.
The modifications represent only 55 source lines of code out of 131,486 for the
three packages studied. Moreover most code modifications are to express the
actual initialization policy, which means existing code can be proven safe. Only
45 methods out of 3,859 (1.1\%) and 2 fields out of 1,524 were annotated. Last
but not least, the execution of the type checker takes less than 20 seconds for
the packages studied.
\remarques{(done) The presentation of information in Table 1 is bulk. I suggest
  it should be visually improved. }

\begin{figure}[t!]\centering
   \includegraphics[scale=0.17]{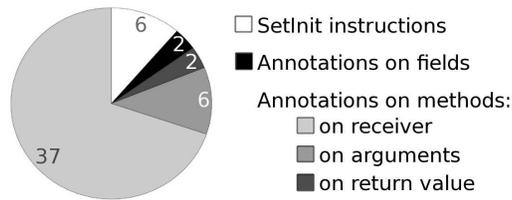}
   \caption{\label{tab:annot-instr} Distribution of the 47 annotations
     and 6 instructions added to successfully type the three packages
     of the JRE.}
   \aftertablecaption{}
 \end{figure}


\paragraph{Adapting the security policy}
Fig.~\ref{tab:annot-instr} details the annotations and the \lstinline!SetInit!
added to specify the security policy.  In the runtime library, a usual pattern
consists in calling methods that initialize fields during construction of the
object. In that case,
a simple annotation \lstinline!@Pre(@Raw(super(C)))! on methods of
class \lstinline!C! is necessary.
These cases represent the majority of the 37 annotations on method receivers.
6 annotations on method arguments are used, notably for some methods of
\lstinline!java.lang.SecurityManager! which check permissions on an object
during its initialization.
The instruction \lstinline!SetInit! is used when a constructor initializes all
the fields of the receiver and then call methods on the receiver that are not
part of the initialization.  In that case the method called need at least a
\lstinline!Raw(C)! level of initialization and the \lstinline!SetInit!
instruction allows to express that the constructor finished the minimum
initialization of the receiver.  Only 6 \lstinline!SetInit! intructions are
necessary.

\paragraph{Cast instructions}
Such a static and modular type checking introduces some necessary loss of
precision ---~which cannot be completely avoided because of computability
issues.  To be able to use our type system on legacy code without deep
modifications, we introduce two dynamic cast operators: \texttt{\small(Init)}
and \texttt{\small(Raw)}.  The instruction \lstinline!y = (Init)x;! allows to
dynamically check that \lstinline!x! points to a fully initialized object: if
the object is fully initialized, then this is a simple assignation to
\lstinline!y!, otherwise it throws an exception.  As explained in
Section~\ref{sec:overview}, the invariant needed is often weaker and the
correctness of a method may only need a $\Raw(c)$ reference.
\lstinline!y = (Raw(C))x! dynamically checks that \lstinline!x! points to an
object which is initialized up to the constructor of class \lstinline!C!.

Only 4 cast instructions are necessary.  There are needed in two particular
cases.  First, when a field must be annotated, but annotation on fields were
only necessary on two fields ---~they imply the use of 3 \lstinline!(Init)! cast
instructions. The second case is on a receiver in a \lstinline!finalize()!
method that checks that some fields are initialized, thereby checking that the
object was \lstinline!Raw(C)! but the type system could not infer this
information.
The later case implies to use the unique \lstinline!(Raw(C))!  instruction
added.

\paragraph{Remaining classes}
Finally, only 4 classes are not well-typed after the previous modifications.
Indeed the compiler generates some code to compile inner classes and part of
this code needs annotations in 3 classes.  These cases could be handled by doing
significant changes on the code, by adding new annotations dedicated to inner
classes or by annotating directly the bytecode.  The one class remaining is not
typable because of the limited precision of our analysis on arrays: one can only
store \texttt{@Init} values in arrays.  To check this later class, our type
system needs to be extended to handle arrays more precisely but this is left for
future work.

\paragraph{Special case of finalize methods}
\remarques{p16, the final discussion before the conclusion is unclear; did you
  find any error in such code?}%
As previously exposed, \lstinline!finalize()!  methods
may be invoked on a completely uninitialized receiver.  Therefore, we study the
case of \lstinline!finalize()!  methods in the packages \lstinline!java.*! and
\lstinline!javax.*!.  In the classes of those packages there are 28
\lstinline!finalize()!  methods and only 12 succeed to be well-typed with our
default annotation values.  These are either empty or do not use their receiver
at all. For the last 16 classes,\remarques{\footnotesize Could you informally
  discuss the impact of what is left outside the formal language,
  e.g. concurrency, and external calls?}  \remarques{\scriptsize Also, how
  difficult would it be to integrate your type system with the base one?  One
  can hardly typecheck code independently for each error pattern.}  the
necessary modifications are either the use of cast instructions when the code's
logic guarantees the success of cast, or the addition of \lstinline!@Pre(@Raw)!
annotations on methods called on the receiver.  In that case, it is important to
verify that the code of any called method is defensive enough.  Therefore, the
type system forced us to pay attention to the cases that could lead to security
breaches or crashes at run time for \lstinline!finalize()! methods.  After a
meticulous checking of the code we added the necessary annotations and cast
instructions that allowed to verify the 28 classes.

\section{Conclusion and Future Work}
\label{sec:conclusion}

We have proposed herein a solution to enforce a secure initialization of objects
in Java. The solution is composed of a modular type system which allows to
manage uninitialized objects safely when necessary, and of a modular type
checker which can be integrated into the BCV to statically check a program at
load time.  The type system has been formalized and proved sound, and the
type-checker prototype has been experimentally validated on more than 300
classes of the Java runtime library.  \remarques{\scriptsize "rare cases
  necessitate .... our solutions for arrays..."  As in Section 4, you're not
  very clear about these "rare cases", and which categories there are.  Most of
  these can be handled using these explicit casts (right?), except a few where
  there are inner classes that need such casts, and then only these rarer cases
  with problematic arrays remains, (right?).  The status of your "solution for
  arrays" is very unclear; in Section 4 you don't even mention that you have
  one, albeit unproven \& unimplemented. I guess you have checked whether it
  would work on the single problematic case in the JRE, right? How
  simple/complex an extension of the system does it involve?}

The experimental results point out that our default annotations minimize the
user intervention needed to type a program and allows to focus on the few
classes where the security policy needs to be stated explicitly. The possible
adaptation of the security policy on critical cases allows to easily prevent
security breaches and can, in addition, ensure some finer initialization
properties whose violation could lead the program to crash.  On one hand,
results show that such a static and modular type checking allows to prove in an
efficient way the absence of bugs. On the other hand, rare cases necessitate the
introduction of dynamic features and analysis to be extended to analyze more
precisely arrays.  With such an extension, the checker would be able to prove
more classes correct, but this is left for future work.

On the formalization side, an obvious extension is to establish the
soundness of the approach in presence of multi-threading. We
conjecture the soundness result still holds with respect to the
Java Memory Model because of the flow insensitive abstraction made
on the heap.  

The prototype and the Coq formalization and proofs can be downloaded from
\url{http://www.irisa.fr/celtique/ext/rawtypes/}.






 \subsubsection*{Acknowledgment.}
 This work was partly supported by the Région Bretagne and by the ANSSI (JavaSec project,
see \texttt{http://www.ssi.gouv.fr/site\_article226.html}).

\bibliographystyle{plain}
\def\tm{\leavevmode\hbox{$\rm {}^{TM}$}}
\bibliography{biblio}

\end{document}